%% file: main.tex
\documentclass{wqcd03}                 

\usepackage{amsmath,amssymb}
\usepackage{graphicx}
\usepackage{dcolumn}
\usepackage{bm}

\newcommand{\bfm}[1]{\mbox{\boldmath$#1$}}

\input{def}
\usepackage{txfonts}                   
\usepackage{mathptmx}                 
                                       
\begin{document}                                       
\confname{QCD@Work 2003 - International Workshop on QCD, Conversano, Italy, 
14--18 June 2003}

\title{Effect of pion thermal width on the sigma spectrum}  
\author{Tetsuo~Nishikawa\addressmark{a}\thanks{Speaker at the Workshop.}, Yoshimasa~Hidaka\addressmark{a}, Munehisa~Ohtani \addressmark{b} and  Osamu~Morimatsu\addressmark{a}}  \address[a]{Institute of Particle and Nuclear Studies,  High Energy Accelerator Research Organization, 1-1, Ooho,  Tsukuba, Ibaraki, 305-0801, Japan} 
\address[b]{The Institute of Physical and Chemical Reseach ( RIKEN ),  2-1 Hirosawa, Wako, Saitama 351-0198, Japan}  

\begin{abstract}
We study the effect of the thermal width of $\pi$ on the spectral function of $\sigma$.
In order to take into account a finite thermal width of $\pi$,
we replace the internal pion mass in the self-energy of $\sigma$ with that of the complex pole found in a previous paper.
The obtained spectral function for $T\aplg 100\,{\rm MeV}$ turns out to possess two broad peaks.
Although a sharp peak at $\sigma\rightarrow\pi\pi$ threshold was observed in the one-loop calculation without the pion thermal width, the peak is shown to be smeared out.
We also search for the poles of the $\sigma$ propagator
and analyze the behavior of the spectral function with these poles.
\end{abstract}
\maketitle

Restoration of the chiral symmetry in hot matter has attracted
much attention since drastic spectral change of light mesons
is expected to be a precursor of the restoration \cite{HK}.
Physical mechanism of the spectral change is explained with the 
mass shift of mesons as follows.
As the temperature increases to the critical temperature,
the mass of $\sigma$ reduces while light $\pi$ becomes heavy,
because their masses should degenerate after the symmetry restoration.
At a certain temperature, therefore, the mass of $\sigma$ coincides with 
twice that of $\pi$. Accordingly, the spectrum of $\sigma$
is expected to be enhanced near the threshold of $\sigma\rightarrow\pi\pi$,
since the phase space available for the decay is squeezed to zero.

Chiku and Hatsuda \cite{chiku} showed that the threshold enhancement in the $\sigma$ channel
is observed as expected.
They also calculated the spectral function of $\pi$ and found that $\pi$ has a finite width 
due to pion absorption in the heat bath:
$\pi+\pi^{\rm thermal}\rightarrow \sigma$.
The width of $\pi$ is $\sim 50\,{\rm MeV}$ at the temperature where the threshold in the $\sigma$ spectrum is most strongly enhanced.
In their analysis of the spectrum of $\sigma$, however,
the effect of the pion thermal width is not included,
since they calculated the self-energy only up to the one-loop order.
The purpose of this paper is to show that the threshold enhancement is smeared out by taking into account the effect of the pion thermal width.

Our strategy is as follows.
We utilize the one-loop self-energy for $\sigma$,
but we replace the masses of internal pions with complex ones.
This complex mass was obtained from the location of the pole of the pion propagator
in a previous work \cite{hidaka}.
Using the self-energy with the complex pion mass, 
we study the spectral function of $\sigma$.
The poles of the $\sigma$ propagator are also searched for
to analyze the behavior of the spectral function.

Along the lines mentioned above, here we employ
the ${\cal O}(4)$ linear sigma model.
It is known that naive perturbation theory breaks down at $T\neq 0$, 
and resummation of higher orders is necessary.
We adopt here a resummation technique called
optimized perturbation theory (OPT) \cite{chiku}.
In OPT one adds and subtracts a new mass term with the mass $m$ 
to the Lagrangian.
One treats the added one as a tree-level mass term while the subtracted one 
as perturbation.
The arbitrary parameter $m$ is determined so that the correction terms
in the pion self-energy are as small as possible \cite{chiku}.

The effects of  
the pion thermal width on the spectrum of $\sigma$
are included in the self-energy at the two or higher loop level.
The two-loop self-energy gives vertex corrections,
mass shifts and thermal widths of $\sigma$ and $\pi$.
Among them, the pion thermal width is the most important 
as an effect beyond the one-loop calculation, from the reason explained in the next paragraph.

\begin{figure}[b]
\begin{eqnarray*}
\parbox{.1\linewidth}{\includegraphics[width=2.5 \linewidth]{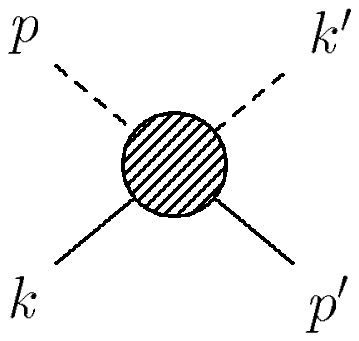}}
\hspace{1.2cm}=&&\hspace{.5cm}\parbox{.1\linewidth}{\includegraphics[width=2\linewidth]{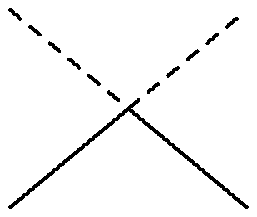}} 
\hspace{1.2cm}+\hspace{.5cm} \parbox{.1\linewidth}{\includegraphics[width=2\linewidth]{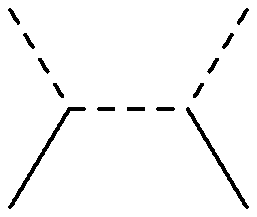}}
\hspace{.5cm}\cr
\\
&&+\hspace{.3cm} \parbox{.1\linewidth}{\includegraphics[width=2\linewidth]{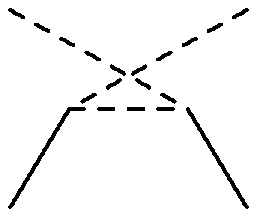}}
\hspace{1.2cm}+\hspace{.5cm} \parbox{.1\linewidth}{\includegraphics[width=2\linewidth]{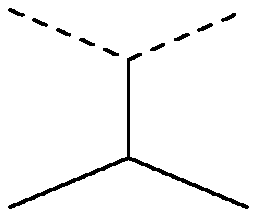}}
\end{eqnarray*}
\caption{The scattering amplitude of $\sigma\pi
\rightarrow\sigma\pi$ at lowest order.}
\label{scatt}
\end{figure}

\begin{figure}[b]
\includegraphics[keepaspectratio,width=7cm]{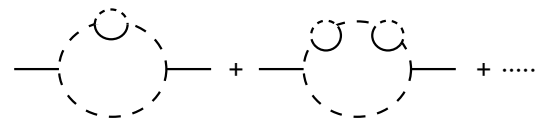}
\caption{The higher-loop self-energies responsible for the thermal width of $\pi$ from the decay $\sigma\rightarrow\pi\pi$.}
\label{scatt-correc}
\end{figure}

Although several processes contribute to the two-loop self-energy,
the most relevant process is the
pion scattering, $\sigma\pi^{\rm thermal}\to\sigma\pi$,  
because $\pi$ is the lightest in the
heat bath and we concentrate on the energy region
near the $\pi\pi$ threshold.
This process contributes to the imaginary part of the two-loop self-energy
in the following way;
\bey
&&\textrm{Im}\Pi^{2-\scriptsize\textrm{loop}}_\sigma(k,T)\sim \int \frac{\textrm{d}^3 p}{(2\pi)^3}\frac{1}{2\omega_{\pi}}\frac{\textrm{d}^3 p'}{(2\pi)^3}\frac{1}{2\omega_{\sigma}}\frac{\textrm{d}^3 k'}{(2\pi)^3}\frac{1}{2\omega_{\pi}}\cr
&&\qquad\times(2\pi)^4\delta^{(4)}(p+k-p'-k')|{\cal M_{\sigma\pi\rightarrow\sigma\pi}}|^2\cr
&&\qquad\times\{n_\pi(1+n_{\sigma})(1+n_\pi)-(1+n_{\pi})n_{\sigma}n_{\pi}\},
\eey
where $n_{\phi}$ is a thermal factor and ${\cal M_{\sigma\pi\rightarrow\sigma\pi}}$ the amplitude of
$\sigma\pi^{\scriptsize\textrm{thermal}}\rightarrow\sigma\pi$.
The tree-level amplitude of the scattering
is shown in Fig.\ref{scatt} and written as
\begin{eqnarray}
{\cal M_{\sigma\pi\rightarrow\sigma\pi}}&=&\frac{-i\lambda}{3}+\frac{-\lambda^2\xi^2}{9}\left[\frac{i}{(k+p)^2-m_{0\pi}^2}\right.\cr
&+&\left.
\frac{i}{(k-p')^2-m_{0\pi}^2}+\frac{3i}{(k-k')^2-m_{0\sigma}^2}\right],
\label{scattamp}
\end{eqnarray}
where $m_{0\phi}$ is the tree level mass.
If $k^2=m_{0\sigma}^2$, $\bm{p},\bm{k}\rightarrow0$, and $m_{0\sigma}\gg m_{0\pi}$, 
four terms in Eq.~(10) cancel out due to the chiral symmetry.
However we are interested in the spectral function $\rho_\sigma(k,T)$ for $k_{0}\equiv\omega\simeq 2m_{0\pi}$, and therefore the third term is dominant; 
\begin{eqnarray}
{\cal M_{\sigma\pi\rightarrow\sigma\pi}}&=&-i\frac{\lambda}{3}\left\{1+(m_{0\sigma}^2-m_{0\pi}^2)\left[\frac{1}{\omega(\omega+2m_{0\pi})}
\right.\right.\cr
&&\left.\left.
+\frac{1}{\omega(\omega-2m_{0\pi})}+\frac{3}{\omega(\omega-2m_{0\sigma})}\right]\right\}\cr
&\simeq&-i\frac{\lambda}{3}\frac{1}{\omega(\omega-2m_{0\pi})}.
\end{eqnarray}
This term comes from the cut diagram shown in Fig.\ref{scatt-correc}. 
This fact indicates that a correction for the self-energy of $\pi$
is the most important contribution for the sigma spectrum beyond the one-loop
level. 
Higher order corrections as shown in Fig.\ref{scatt-correc}
are also large.
Therefore, it is necessary to resum at least higher order corrections of this class of diagrams. 
This resummation can be performed by replacing the internal pion propagators of $\Pi_\sigma$ with the pion propagators in the one-loop level.
Instead of using the complicated pion propagator in the one-loop level, we adopt  here a simplified prescription --- to use a complex pion pole
as internal pion mass in $\Pi_{\sigma}$, in order to include the most important effect.
The thermal width of $\pi$ 
is represented as the imaginary part of 
the complex pole of the pion propagator:
\bey
&&\left. p^2-m_{0\pi}^2
-\Pi_\pi(p^2, {\bfm{p}}\; ; T) \right|_{p^2=(m_\pi^{\rm pole})^2}=0, \\
&&m_\pi^{\rm pole}=m_\pi^{*}(T)-i \frac{\varGamma_\pi(T)}{2},
\eey
where $\Pi_{\pi}$ is the one-loop pion self-energy \cite{hidaka}.

\begin{figure}
\begin{center}
\includegraphics[keepaspectratio,height=3cm]{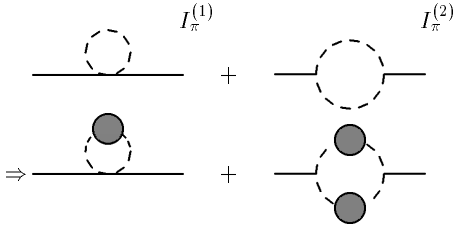}
\caption{A Prescription to take into account the thermal width of $\pi$ in the decay $\sigma\rightarrow\pi\pi$. The upper diagrams represent the self-energy of with the internal pion mass, $m_{0\pi}$, and the lower diagrams with $m_{\pi}^{\rm pole}$.}
\label{diagram}
\end{center}
\end{figure}

As a matter of course, the pole has a dependence 
on $|\boldsymbol{p}|$ naturally \cite{hidaka},
but contributions from $|\boldsymbol{p}|\gtrsim T$ 
are less effective to the loop integration in the thermal part of $\Pi_\sigma$.
Therefore, as a simple prescription to incorporate 
the thermal width,
we replace the internal pion mass, $m_{0\pi}$, in the one-loop self-energy with $m_\pi^{\rm pole}$ at $\bm{p}={\bf 0}$ 
in Ref.\cite{hidaka} as depicted in Fig.\ref{diagram} 
to obtain the spectral function $\rho_\sigma$.

\begin{figure}
\begin{center}
\includegraphics[width=8.5cm]{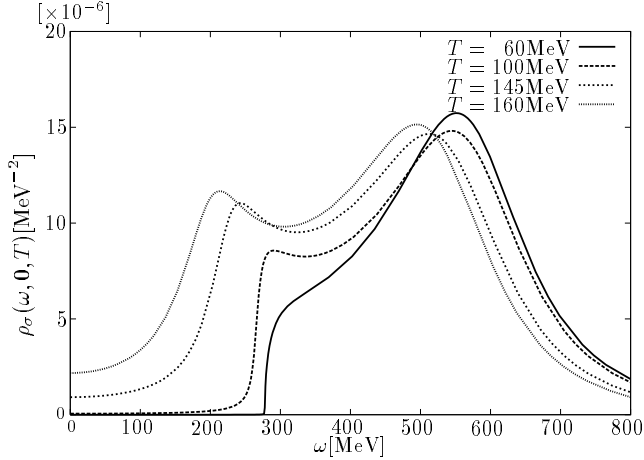}
\caption{Spectral function of $\sigma$ with the thermal width of $\pi$.}
\label{spectral}
\end{center}
\end{figure}

\begin{figure}
\begin{center}
\includegraphics[width=8.5cm]{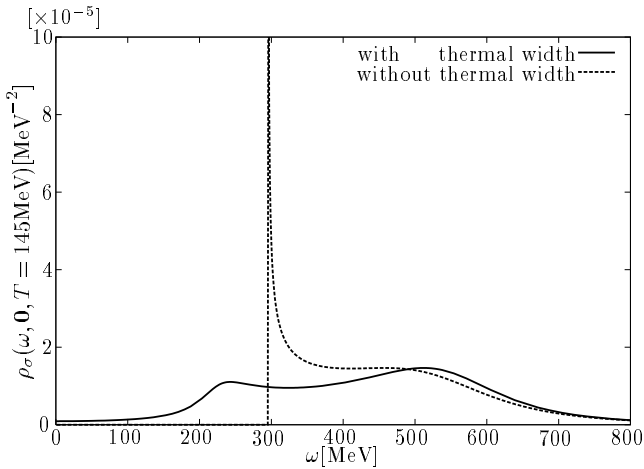}
\caption{Spectral function of $\sigma$ with and without the thermal width of $\pi$ at $T=145\,{\rm MeV}$.}
\label{hikaku}
\end{center}
\end{figure}

We show in Fig.\ref{spectral} the obtained spectral function
at $\bm{k}={\bf 0}$ for several values of $T$.
For comparison, we also show the spectral function at $T=145\,{\rm MeV}$
with $m_{0\pi}$ and that with $m_\pi^{\rm pole}$ in Fig.\ref{hikaku}.
At low $T$ the spectrum consists of a broad bump 
around $\omega=550\,{\rm MeV}$.
As $T$ increases, another strength grows
at the left shoulder of the bump,
while the spectrum for $\omega\aplg 400\,{\rm MeV}$ does not vary much.
In contrast to this case, 
if $m_{0\pi}$ is used as the internal pion mass,
the spectrum near the threshold of $\sigma\rightarrow\pi\pi$ is enhanced as $T$ is increased. 
At $T=145\,{\rm MeV}$, where the mass of $\sigma$ coincides with $2m_{0\pi}$,
the spectrum near the threshold
is most strongly enhanced as shown in Fig.\ref{hikaku} (dotted curve).
When $m_{\pi}^{\rm pole}$ is used, however, 
due to the thermal width of $\pi$ in the decay $\sigma\rightarrow\pi\pi$,
the sharp threshold is smeared to be a bump at $\omega=200\,{\rm MeV}\sim 300\,{\rm MeV}$ as the solid curve in Fig.\ref{hikaku}.

In general, the behavior of the spectral function is governed 
by poles of the propagator.
We thus next search the complex $\omega$ plane for the poles of the propagator.
The analytic structure of the propagator is determined by that of the self-energy.
Then, we see that the self-energy has a branch point at $\omega=2m_\pi^{\rm pole}$ around which the Riemann sheet is two-fold.
The self-energy has another branch point due to the $\sigma$ loop which, however, is irrelevant for the present discussion of the self-energy around the threshold of $\sigma\rightarrow\pi\pi$.
This is because the branch point risen from the $\sigma$ loop 
is located at the twice of the $\sigma$ mass, which is  
far from the $\pi\pi$ threshold even if $\rho_\sigma$
has some strength below the threshold.
The branch cut from $\omega=2m_\pi^{\rm pole}$ can be arbitrarily chosen at finite temperature, because it has no physical meaning.
For definiteness we choose the cut from the branch point straight to the right, parallel to the real axis and refer to the sheet including the physical real energy as the first Riemann sheet and the other one as the second Riemann sheet.
We show in Fig.{\ref{poleNZwidth} the result of searching the locations of the poles on the Riemann sheets.
\begin{figure}
\begin{center}
\includegraphics[width=8.5cm]{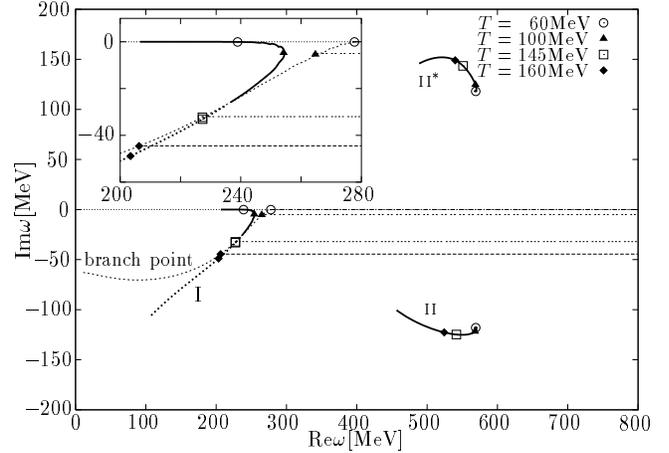}
\caption{Location of the poles of the $\sigma$ propagator with the thermal width of $\pi$,
$m_{\sigma}^{\rm pole}$.
The branch point is shown by dotted line and the branch cut are also indicated.
We enlarge the figure near the branch point and show it 
in the inserted box.}
\label{poleNZwidth}
\end{center}
\end{figure}

We found three relevant poles.
Two of them, (II) and (II*), are on the second Riemann sheet
and off the real axis. Their positions are approximately symmetric 
with respect to the real axis with each other.
For the spectral function of $\sigma$, on one hand,
(II) and (II*) show up as a bump around $\omega=550\,{\rm MeV}$ and
the small movement of these poles explains the insensitivity 
to $T$ of the spectrum for $\omega \gtrsim 400\,{\rm MeV}$.
On the other hand, the structure of the spectrum 
for low $\omega$ is mainly determined by
the remaining one pole, (I).
For preparation to consider this pole (I), 
let us first recall 
the movement of 
the corresponding pole in case of no thermal width 
of $\pi$ \cite{hidaka}.

Without the pion width,
the counterpart of the pole (I)
exists on the second Riemann sheet apart from the branch point of
$2m_{0\pi}$ for $T \lesssim 60\,{\rm MeV}$.
Correspondingly, the $\sigma$ spectrum has no peak
at the $\pi\pi$ threshold for low $T$.
As $T$ increases, the pole moves along the real axis
from below toward the branch point $2m_{0\pi}$ and turns around 
at the point to appear on the first sheet at $T=145\, {\rm MeV}$.
In response to this movement, the $\sigma$ spectrum
is strongly enhanced for $T \sim 145 \,{\rm MeV}$.
In other words, the threshold enhancement 
originates in the fact that the $\sigma$ pole approaches 
the branch point on the real axis near the temperature concerned.

Keeping this in mind, 
we return to the discussion on movement of 
the pole (I) next. 
 As is the case with no pion width,
this pole also resides on the second Riemann sheet for low $T$
and comes closer to the branch point with increasing $T$ 
as shown in Fig.{\ref{poleNZwidth}.
At $T=138\,{\rm MeV}$, the pole crosses the branch cut and appears
on the first Riemann sheet from the second, likewise.
%
 The branch point, however,
is located at $2m_\pi^{\rm pole}$ apart from the real axis
for $T\gtrsim 100 \, {\rm MeV}$ due to the finite thermal width of $\pi$ 
(see Fig.{\ref{poleNZwidth}).
The complex branch point demands that 
the pole (I), whose counterpart caused the threshold enhancement
by moving toward the point, 
also acquires an imaginary part near $T \sim 138\,{\rm MeV}$.
As a result, the $\sigma$ spectrum is smeared and
the remarkable contrast is brought about
for the spectrums as shown in Fig.\ref{hikaku}.

Let us argue on the correlation between the spectral function and poles 
of the propagator more quantitatively.
For this purpose,
we approximate the propagator of $\sigma$ with
the superposition of the three pole contributions in the following way:
\bey
&&\rho_\sigma \simeq -2{\rm Im} \sum_{\rm pole} \frac{Z^{\rm pole}}{q-q_{\rm pole}},\cr
&&
Z^{\rm pole} = \frac{1}{2q_{\rm pole}}
\left(  1-\frac{\partial\Pi_\sigma}{\partial k^2}
 (k^2=(m_\sigma^{\rm pole})^2)\right)^{-1},
\label{approx-rho}
\eey
where $m_\sigma^{\rm pole}$ stands for the pole of the $\sigma$ propagator.
$q$ and $q_{\rm pole}$ are defined by $q=(k^2-(2m_\pi^{\rm pole})^2)^{1/2}$
and $q_{\rm pole}=((m_{\sigma}^{\rm pole})^2-(2m_\pi^{\rm pole})^2)^{1/2}$, respectively.
It should be noted that we have taken not $k^2$
but $q=(k^2-(2m_\pi^{\rm pole})^2)^{1/2}$ as an appropriate variable.
The use of this variable enables us to
unfold two leaves of the Riemann sheet around the branch point \cite{newton};
We can reflect which Riemann sheet accommodates 
each pole on the approximated propagator.
We show the approximated spectral function, \eq{approx-rho}, 
for one case, $T=145\,{\rm MeV}$, in \ref{polefit145}.
\begin{figure}[t]
\begin{center}
\includegraphics[width=8.5cm]{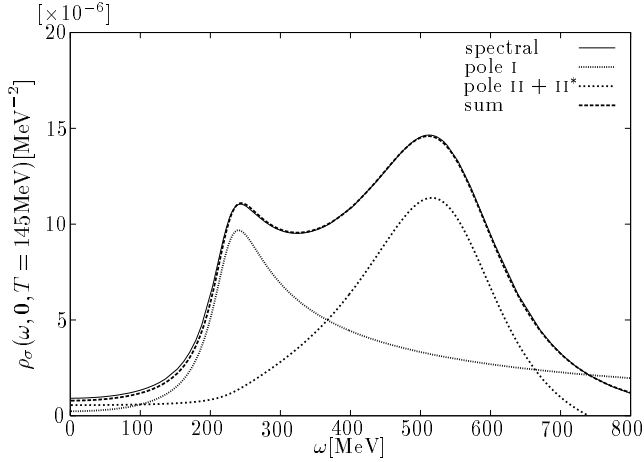}
\caption{The spectral function approximated by a superposition of the contributions from the three poles, \eq{approx-rho}, at $T=145\,{\rm MeV}$.}
\label{polefit145}
\end{center}
\end{figure}
At $T=60\,{\rm MeV}$, the pole (I) is located on the second Riemann sheet,
while it is on the first sheet at $T=145\,{\rm MeV}$.
We can see that the poles (II) and (II*) provide the broad peak around $\omega=550\,{\rm MeV}$ and that the pole (I) mainly determines 
the behavior of the left shoulder of the peak.
If we include only the contributions of (I) and (II), 
two broad peaks are still reproduced.
However, a structure which is not seen in the spectral function appears at low $\omega$ region.
The contribution of (II*) cancels out the spurious structure caused by (II).  
At $T=60\,{\rm MeV}$, the left shoulder of the peak is small,
although the pole (I) seems to lie close to the real axis.
This is because the pole is on the second Riemann sheet 
and is practically far from the real axis.
In contrast, at $T=145\,{\rm MeV}$, the pole (I) is on the first Riemann sheet
and is close to the real axis.
Accordingly, it provides a peak with more strength around $\omega=200\sim300\,{\rm MeV}$.
Thus, in order to reveal the relation 
between the poles and the behavior of the spectrum,
it is necessary to know both of the complex structure of the Riemann sheet 
and the positions of the poles on it.

Before concluding this paper, a comment on the complex-pole approximation is in order.
As, we have shown above, the behavior of the pole of the sigma propagator in the complex energy plane, which determines the smearing of the sigma threshold strength, is essentially governed by the branch point of the sigma self-energy.
And the branch point of the self-energy is correctly given by the complex-pole approximation.
Therefore, the results of the present paper are expected to be basically intact beyond the complex-pole approximation. 

In conclusion, the thermal width of $\pi$ smears out the sharp peak of the $\sigma$ spectrum at $\pi\pi$ threshold.
The structure of the spectral function
can be well understood from the three relevant poles of the propagators 
and the branch points of the self-energy.
Physically we can interpret the origin of the smearing as follows.
By replacing the masses of $\pi$ in the decay $\sigma\rightarrow\pi\pi$ with the complex mass,
processes like $\sigma\pi^{\rm thermal} \rightarrow\sigma\pi$
are newly taken into account.
Since this process is 
allowed for any energy of the initial $\sigma$ \cite{nishi},
$\pi\pi$ threshold disappears thereby.
Due to the above process, the $\sigma$ peak at the threshold is expected to
acquire a width of 
\bey
\varGamma_{\sigma}\sim n_\pi\cdot\sigma_{\pi\sigma}\cdot m_{\pi}^3
\sim 100\,{\rm MeV}
\label{ordergamma}
\eey
for $T\sim140\,{\rm MeV}$,
where $E_\pi$ is the typical energy of thermal pions.
$\sigma_{\pi\sigma}$ is the scattering cross section of $\sigma$ with $\pi$ and its value 
is estimated to be ${\cal O}(1/f_{\pi}^2)$ from the low energy theorem.
The calculated spectral function using the complex pion mass is consistent with the estimate, \eq{ordergamma}.

The smeared behavior of $\rho_\sigma$ is also seen for
finite density \cite{chanf,chaos},
though the physical origin of the width is different.
It is interesting to apply the prescription adopted here 
in the context of finite density systems.

What we have observed in this paper is expected to be universal for processes in which a particle decays into unstable particles.

\def\Ref#1{[\ref{#1}]} 
\def\Refs#1#2{[\ref{#1},\ref{#2}]} 
\def\npb#1#2#3{{Nucl. Phys.\,}{\bf B{#1}},\,(#3)\,#2} 
\def\npa#1#2#3{{Nucl. Phys.\,}{\bf A{#1}},\,(#3)\,#2} 
\def\np#1#2#3{{Nucl. Phys.\,}{\bf{#1}},\,(#3)\,#2} 
\def\plb#1#2#3{{Phys. Lett.\,}{\bf B{#1}},\,(#3)\,#2} 
\def\prl#1#2#3{{Phys. Rev. Lett.\,}{\bf{#1}},\,(#3)\,#2} \def\prd#1#2#3{{Phys. Rev.\,}{\bf D{#1}},\,(#3)\,#2} 
\def\etal{{\it et al.}}

\end{document}

%% file: def.tex
\def\ben{\begin{equation}}
\def\een{\end{equation}}
\def\bey{\begin{eqnarray}}
\def\eey{\end{eqnarray}}
\def\ba{\begin{array}}
\def\ea{\end{array}}
\def\benmrt{\begin{enumerate}}
\def\eenmrt{\end{enumerate}}

\def\psla{p{\raise1pt\hbox{$\!\!/$}}}
\def\dsla{\partial{\raise1pt\hbox{$\!\!\!/$}}}
\def\Dsla{D{\raise1pt\hbox{$\!\!\!/$}}}
\def\xsla{x{\raise1pt\hbox{$\!\!\!/$}}}

\def\jmu5{j_{\mu 5}^{(i)}(0)}
\def\jnu5{j_{\nu 5}^{(i)}(0)}

\def\qq0v{\langle0\!\mid\!{\bar q}q\!\mid\! 0\rangle}

\def\qc0f{\langle0\!\mid\!{\bar q}q\!\mid\!0\rangle_{F}}

\def\qsq0f{\langle0\!\mid\!{\bar q}\sigma_{\mu\nu}q\!\mid\!0\rangle_{F}}

\def\qgdq0f{\langle0\!\mid\!{\bar q}{\cal 
S}\gamma_{\mu}D_{\nu}q\!\mid\!0\rangle_{F}}

\def\qddq0f{\langle0\!\mid\!{\bar q}{\cal
S}D_{\mu}D_{\nu}q\!\mid\!0\rangle_{F}}

\def\3mmtm{|{\bf q}|^2}

\def\eq#1{Eq.(\ref{#1})}

\def\Ref#1{[\ref{#1}]}
\def\Refs#1#2{[\ref{#1},\ref{#2}]}

\def\p0{p_0}

\def\gam3{\mbox{\boldmath{$\gamma$}}}
\def\e0{E_{0}(s_{0},s)}
\def\e1{E_{1}(s_{0},s)}
\def\e2{E_{2}(s_{0},s)}

\def\aplt{\kern0.3333em \raise 0.2ex \hbox{$<$}%
\kern-0.8em \lower0.8ex \hbox{$\sim$}%
\kern0.3333em}
\def\aplg{\kern0.3333em \raise 0.2ex \hbox{$>$}%
\kern-0.8em \lower0.8ex \hbox{$\sim$}%
\kern0.3333em}




%% file: main.bbl
\begin{thebibliography}{99}
\bibitem{HK} T.\ Hatsuda and T.\ Kunihiro, \prl{55}{158}{1985}; \plb{185}{304}{1987}.
\bibitem{chiku}
S.\  Chiku and T.\ Hatsuda, \prd{57}{R6}{1998}, \prd{58}{76001}{1998}.
\bibitem{hidaka} 
Y.\ Hidaka, O.\ Morimatsu and T.\ Nishikawa, \prd{67}{056004}{2003}.
\bibitem{stev} 
P.\ M.\ Stevenson, \prd{23}{2916}{1981}.
\bibitem{newton}
R.\ G.\ Newton, {\it SCATTERING THEORY OF WAVES AND PARTICLES}
(New York, USA: Springer, 1982).
\bibitem{nishi}
T.\ Nishikawa, O.\ Morimatsu and Y.\ Hidaka, {\tt hep-ph/0302098}.
\bibitem{chanf}
G.\ Chanfray, {\tt nuc-th/0212085}.
\bibitem{chaos}
F.\ Bonutti \etal, CHAOS Collaboration, \npa{677}{213}{2000}; \prl{89}{222302}{2002}.
\end{thebibliography}
